\documentstyle[12pt]{article}
\title{
Spectral Flow in  Josephson Junctions and Effective Magnus Force}
\author{ Yu.G. Makhlin and G.E. Volovik\\
Low Temperature Laboratory\\
Helsinki University of Technology\\
Otakaari 3A, 02150 Espoo, Finland\\
and\\
L.D. Landau Institute for Theoretical Physics\\
Kosygin St. 2, 117940 Moscow, Russia\\
}
\begin{document}
\maketitle
\begin{abstract}
{ Momentum production during the phase slip process in SNS
Josephson junction is discussed. It is caused by the spectral
flow of bound states of fermions localized within the junction.
This effectively reduces the Magnus force acting on vortices
which provides an explanation for the experimental observation
of the negligible Magnus force in 2D Josephson junction arrays.
The flow of the fermionic levels is similar to that in sphalerons
in particle physics, where it gives rise to the baryogenesis. }
\end{abstract}

\vfill \eject

Vortex dynamics in 2D Josephson junction arrays has been studied
intensively in recent years. The vortices in the arrays could be
considered as massive particles with long-range Coulomb
interaction \cite{EckernSchmid,FazioSchoen}. A particular
attention was devoted to the forces acting on the vortices, one
of them being Magnus force.

In the experiment \cite{ballistic} the straightforward ballistic
motion of vortices was observed which implies the absence of
reactive forces acting on a vortex perpendicular to its
velocity. This was also confirmed by more recent experiments:
vortices move perpendicular to the driving current
\cite{Lachenmann}, and no Hall effect was detected in the system
\cite{noHall}.

An explanation of the absence of the Magnus force was proposed
in Ref.\cite{SchoenMagnus}. The authors claim that the Magnus
force is proportional to the offset charges on the
superconducting islands, the effect of which being negligible.
In a recent worked \cite{GaitanJJA} Gaitan and Shenoy argued
that the Magnus force is proportional to the density of
superconducting electrons on the islands averaged over distances
large compared to the lattice constant of the array rather than
to the charge of the island which is given by the difference in
the numbers of electrons and protons. On the other hand Zhu, Tan
and Ao \cite{ZhuTanAo} have shown the force to be proportional
to the {\it local} superconducting density at the point where
the vortex is situated. Since the vortex does not move through
superconducting islands but through the junctions, the Magnus
force on the vortex can be substantially reduced.

In the present paper we consider the forces on a vortex moving
in a Josephson junction array, and propose a different
explanation to the experiments mentioned above. It should be
emphasized that one contribution to the force was missed in
previous considerations \cite{SchoenMagnus,GaitanJJA,ZhuTanAo}
which is the force from the {\it spectral flow}. In uniform
superfluids there exists the so called zero branch of energy
levels of fermions localized within vortex core which crosses
zero as a function of angular momentum \cite{Caroli}. Vortex
motion leads to the flow of the levels along this branch, and
energy of some levels crosses zero value. At low temperature
these levels become occupied (or unoccupied, depending on
whether they cross zero upwards or downwards). During this
process the whole number of quasiparticles localized within the
core is conserved, while the linear momentum of the
quasiparticles is not conserved. This implies a transfer of the
linear momentum from the vortex to the heat bath and thus an
additional force acting on the moving vortex
\cite{CallanHarvey}. This force from the spectral flow can
almost completely neutralize Magnus force
\cite{KopninVolovikParts} though in some regimes the spectral
flow is suppressed and the net force appears. This scenario
reproduces the microscopic calculations by Kopnin and co-authors
\cite{KopninCoAuthors}.

We argue that an analogous situation takes place in Josephson
junction arrays. Here the role of fermions localized within a
vortex core is played by fermions localized within junctions.
The phase slip events in junctions during vortex motion are
accompanied by the spectral flow of the bound state fermions.
The force from this spectral flow neutralizes Magnus force, and
the reactive force on a vortex becomes negligible.

We consider a particular example of a Josephson junction array
and show the cancelation of two contributions to reactive
force. Other geometries and possible difference in Magnus and
spectral flow forces are under investigation.

We consider a square lattice of superconducting islands as
plaquettes, the edges of the lattice representing normal layers
in SNS-junctions. Vortices could be considered as sitting at
vertices of the lattice (Fig.\ref{lattice}).

As in the review \cite{SoninReview} and Ref.\cite{KopninVolovikParts}
(see also \cite{Volovik2}) we
assume that the conventional Magnus force on a vortex at low
temperature is determined by the local density $n$ of the
electron because $n$ is the variable which is canonically
conjugated to the phase of the Bose-condensate:
\begin{equation}
{\bf F}_M=\pi n \hat{\bf z}\times{\bf v}_L~~,
\label{Magnusforce}
\end{equation}
Here ${\bf v}_L$ is the velocity of the vortex with respect to
the superfluid velocity, which is chosen to coincide with the
velocity of the heat bath (${\bf v}_s={\bf v}_n=0)$. The 2D
particle density of electrons $n$ is assumed to be the same in
the superconducting and normal regions. Note that in the
low-temperature limit the superfluid density $n_s$ tends to $n$
irrespective of the magnitude of the order parameter gap
$\Delta$ (actually due to the Iordanskii force from the heat
bath, the Eq.(\ref{Magnusforce}) is valid even at nonzero $T$
\cite{SoninReview,Volovik2}). This can be applied also to the
electrons in ``normal'' regions with small magnitude of
$\Delta$. This expression does not contradict to
Ref.\cite{GaitanJJA} at $T=0$, but we want to demonstrate that
there is another force which nearly completely compensates the
Magnus force.

During one step a vortex moves from a vertex to an adjacent one.
We are going to calculate production of momentum due to spectral
flow of fermions for such a process, which gives rise to the
compensating force. The problem is in many details similar to
the evolution of fermionic bound states, which exist within
topological solitons in polymers, superfluid $^3$He and other
ordered systems in condensed matter with fermions
\cite{JackiwSchriefferHoEtc}. The same type of evolution of the
fermionic spectrum in topological solitons and sphalerons in
particle physics leads to baryogenesis \cite{Turok,Diakonov},
while in our case the spectral flow leads to "momentogenesis".

Let us consider one particular Josephson junction with $x$ being
the coordinate axis normal to the junction (Fig.\ref{lattice}).
The dependence of the eigenfunctions of Bogolyubov-Nambu hamiltonian
\begin{equation}
{\cal H}=\hat\tau_3\cdot\hat\epsilon+\hat\tau_1 {\rm Re}\Delta-
\hat\tau_2 {\rm Im}\Delta
\end{equation}
on ${\bf r}_\perp=(y,z)$ is given by a factor $\exp(i{\bf
k}_\perp {\bf r}_\perp)$. Here $\hat{\vec\tau}$ are Pauli
matrices in Bogolyubov-Nambu space and
$\hat\epsilon=(-\nabla^2-k_F^2)/2m^*$ is the operator of energy
of quasiparticles in normal liquid; $k_F$ is the Fermi momentum.
We suppose that the order parameter varies slow on the length
scale of the coherence length, while the Fermi momentum remains
constant. We are interested in the energies close to Fermi
surface, so, we suppose that $k_\perp<k_F$ and the eigenfunction
is represented in eikonal approximation:
\begin{equation}
\exp(iqx)\cdot \psi(x)
\end{equation}
where $q^2=k_F^2-k_\perp^2$ ($q$ plays a part of Fermi momentum
of 1D Fermi liquid for given ${\bf k}_\perp$), the exponent
represents fast oscillations in space, and $\psi(x)$ varies
slowly. In this case we may substitute $\hat\epsilon$ by
$q(-i\nabla)/m^*$. The approximation gives energy spectrum with
the accuracy small compared to $\Delta$.

We shall investigate the dependence of the energy spectrum on the
phase difference $\Delta\phi$ between two superconducting
islands. For simplicity we suppose that the order parameter is
real $\Delta(x)=|\Delta(x)|$ on the one side of the junction ($x<0$),
and on the other side ($x>0$) the order parameter is given by
$\Delta(x)=|\Delta(x)|\cdot\exp(i\phi)$. For $\phi=\pi$ the order
parameter is an odd real function of $x$ (Fig.\ref{Dofx}) and the
hamiltonian is supersymmetric: $\{{\cal H},\tau_2\}=0$. This
$0-\pi$ soliton corresponds to the sphaleron in particle physics
\cite{Turok,Diakonov}, i.e. the intermediate state between vacua
with different topological charges. Due to supersymmetry the
Hamiltonian has an eigenfunction
\begin{eqnarray}
\tilde\psi_0={\rm const}
\left(\begin{array}{c}1\\ -{\rm sign(q)}i\end{array}\right)\psi_0,\nonumber\\
\psi_0=\exp\left(-\frac{m^*}{|q|} \int\limits_0^x dx^\prime \Delta(x^\prime)
\right).
\end{eqnarray}
with zero eigenvalue. So, at $\phi=\pi$ for each $q$ one energy level
crosses zero. For small $\phi-\pi$ the perturbation of the hamiltonian is
\begin{equation}
{\cal H}_{int}=-(\phi-\pi)\Delta(x)\hat\tau_2
\end{equation}
for $x>0$ and ${\cal H}_{int}=0$ for $x<0$. The energy level is shifted to
\begin{equation}
E={\rm sign}(q)(\phi-\pi)\omega(|q|),  \label{E}
\end{equation}
where
\begin{equation}
\omega(|q|)=\frac{\int\limits_0^\infty dx |\Delta(x)| \psi_0^2(x)}
                 {\int\limits_0^\infty dx \psi_0^2(x)}.
\end{equation}
{}From (\ref{E}) it follows that at $\phi=\pi$, i.e. at the
sphaleron, the energy level crosses zero upwards for $q>0$ and
downwards for $q<0$. The similar phenomenon (cf.
Refs.\cite{Kashiwaya,Tanaka} as well) was found for sphalerons
in particle physics.

This leads to the production of the $x$-component of the linear momentum
\begin{equation}
\Delta P=2\cdot\frac{1}{2} A
\int \frac{d^2k_\perp}{(2\pi)^2}
2|q|=\frac{k_F^3}{3\pi}A.
\end{equation}
The prefactor $2$ stands for double spin degeneracy, and $1/2$
removes the double counting of particle and hole momenta. Here
$A$ is the area of the cross-section of the junction, which is
given by the product of the lattice constant $a$ and the
thickness $h$ of the film along $z$-axis (which is perpendicular
to the plane): $A=ah$.

Using this equation one can find the spectral-flow force
experienced by the vortex moving with respect to the heat bath.
Since the velocity of the vortex is $v_L=a/\Delta t$, where
$\Delta t$ is the period of time during which the vortex crosses
one junction, one has for the spectral-flow force in $x$
direction
\begin{equation}
F_{sp.flow}=\frac{\Delta P}{\Delta t}= \frac{\Delta P v_L}{a}=\pi C_0 v_L,
\end{equation}
where $C_0=h k_F^3/3\pi^2$. In vector notations
\begin{equation}
{\bf F}_{sp.flow}=\pi C_0 {\bf v}_L \times \hat{\bf z}.
\end{equation}

The 2D particle density is $n=\rho h$ where the 3D density $\rho
\approx k_F^3/3\pi^2$ with the accuracy of the order of
$(\Delta/E_F)^2$ . Therefore, the force induced by the spectral
flow nearly neutralizes the Magnus force as anticipated.

{\it Discussion.\/} Let us now make some remarks on a more
general case of the geometry of Josephson junction array when
superconducting islands do not cover almost all area of a
sample. If the core size of a vortex $r_c$ is small compared to
the lattice spacing $a$ then one has the situation discussed in
\cite{ZhuTanAo}: the vortex can be considered as a point-like
object moving in the locally homogeneous environment with the
slowly changing density of the electrons. In this case one can
apply the bulk results \cite{KopninVolovikParts}: the {\it en
route} Magnus force will be canceled by the {\it en route}
spectral-flow force for any route.

Let us consider another extreme limit $r_c\gg a$, which takes
place if superconducting islands cover a small part of the area.
In this case considered in Ref.\cite{GaitanJJA} one can average
over the distances of the order of the core size $r_c$ and
obtains the case of conventional homogeneous superfluid with
$r_c$ as a coherence length and $a$ as ``interatomic distance''. The
Magnus force in this case will be determined by the average
number density $\bar n$ as it was stated in
Ref.\cite{GaitanJJA}, however the spectral flow force on a
vortex will be also proportional to average value of $C_0$. Thus
they again will nearly compensate each other, but now with the
relative accuracy of the order of $(a/r_c)^2$.

In the intermediate case of $r_c\sim a$ the difference in two
contributions to the reactive force can become comparable to
their values, because the {\it en route} values of $n$ and $C_0$
can be different. Note that there are at least 3 different
reasons which can violate the fine tuning between $\bar n$ and
$C_0$: (1) The particle hole asymmetry gives the disbalance
$\bar n - \bar C_0 \sim \bar n (\Delta/E_F)^2$. (2) The inhomogeneity
discussed above gives $\bar n - \bar C_0 \sim \bar n (a/r_c)^2$. (3)
Finite spacing $\omega_0$ between the quasiparticle energy
levels, which supresses spectral flow of fermions:
$\bar n - \bar C_0 \sim
\bar n (\omega_0\tau)^2$ \cite{KopninVolovikParts,{vanOtterlo}}.
In conventional Abrikosov vortices the spacing is $\omega_0\sim
\Delta^2/E_F$, while in the Josephson junction the discreteness
arises due to the dimensional quantization along the axis $y$,
which is parallel to the junction.

{\it Acknowledgments.\/}
We are grateful to G.~Sch\"on and N.B.~Kopnin for valuable
discussions, and to F.~Gaitan for sending us the preprint
\cite{GaitanJJA} and to S.~Kashiwaya for the preprint \cite{Kashiwaya}.
This work was supported through the ROTA
co-operation plan of the Finnish Academy and the Russian Academy
of Sciences. G.E.V. was also supported by the Russian Foundation
for Fundamental Sciences, Grant Nos. 93-02-02687 and
94-02-03121. Yu.G.M. was also supported by the International
Science Foundation and the Russian Government, Grant No.MGI300,
and by the ``Soros Post-Graduate Student'' program of the Open
Society Institute.

\newpage
\pagestyle{empty}

\begin{figure}[h]
\begin{center}

\caption[lattice]{%
Josephson junction array. The squares represent superconducting
islands and white spacings between them are normal intermediate
layers in junctions. The distribution of phases of
superconducting islands for a vortex sitting on a site of the
lattice marked by a black circle is shown by arrows. When the vortex moves
to the site marked by a dashed circle some levels of fermions localized
within the junction between two central squares cross zero,
and some amount of linear momentum $k_x$ is produced.
This corresponds to an additional force on the vortex which
compensates conventional Magnus force.
}
\label{lattice}
\end{center}
\end{figure}

\begin{figure}[h]
\begin{center}
\caption[Dofx]{%
The dependence of superconducting order parameter on the
coordinate $x$ normal to a junction for phase difference of
$\pi$ between superconducting islands. Corresponding
Bogolyubov-Nambu hamiltonian has zero energy eigenvalues,
i.e. at the moment when $\phi$ crosses $\pi$ the levels cross zero.
}
\label{Dofx}
\end{center}
\end{figure}


\begin{thebibliography}{99}
\bibitem{EckernSchmid}
U.~Eckern, A.~Schmid, Phys. Rev. {\bf B 39}, 6441 (1989).

\bibitem{FazioSchoen}
R.~Fazio, G.~Sch\"on, Phys. Rev. {\bf B 43}, 5307 (1991).

\bibitem{ballistic}
H.S.J.~van~der~Zant, F.C.~Frischy, T.P.~Orlando, and J.E.~Mooij,
Europhys. Lett. {\bf 18}, 343 (1992).

\bibitem{Lachenmann}
S.G.~Lachenmann, T.~Doderer, D.~Hoffmann, R.P.~Huebner, P.A.A.~Booi,
and S.P.~Benz, Phys. Rev. {\bf B 50}, 3158 (1994).

\bibitem{noHall}
H.S.J.~van~der~Zant, M.N.~Webster, J.~Romijn, and J.E.~Mooij,
Phys. Rev. {\bf B 74}, 4718 (1995).

\bibitem{SchoenMagnus}
R.~Fazio, A.~van~Otterlo, G.~Sch\"on, H.S.J.~van~der~Zant, J.E.~Mooij,
Helv. Phys. Acta {\bf 65}, 228 (1992).

\bibitem{GaitanJJA}
F.~Gaitan and S.R.~Shenoy, preprint, cond-mat/9505088.

\bibitem{ZhuTanAo}
X.-M.~Zhu, Yong Tan, and P.~Ao, preprint, cond-mat/9507126.

\bibitem{Caroli}
C.~Caroli, P.G.~de~Gennes, and J.~Matricon, Phys. Lett. {\bf
9}, 307  (1964).

\bibitem{CallanHarvey}
G.E.~Volovik, Pis'ma v ZhETF {\bf 57}, 233 (1993)
[JETP Lett. {\bf 57}, 244 (1993)].

\bibitem{KopninVolovikParts}
N.B.~Kopnin, G.E.~Volovik, \"U.~Parts, preprint, cond-mat/9509157.

\bibitem{KopninCoAuthors}  N.B. Kopnin, V.E. Kravtsov,  Pis'ma ZhETF
{\bf 23}, 631 (1976)  \lbrack JETP Lett. {\bf 23}, 578 (1976)\rbrack
N.B. Kopnin, V.E. Kravtsov, ZhETF {\bf 71}, 1644 (1976)  \lbrack  JETP  {\bf
44}, 861 (1976)\rbrack;  N.B. Kopnin, A.V. Lopatin,  Phys. Rev. {\bf B 51},
15291 (1995).

\bibitem{SoninReview}
E.B.~Sonin, Rev. Mod. Phys. {\bf 59}, 87 (1987).

\bibitem{Volovik2}
G.E.~Volovik, Pis'ma ZhETF {\bf 62}, 58 (1995).

\bibitem{JackiwSchriefferHoEtc} R. Jackiw, and J.R. Schrieffer, Nucl.
Phys. {\bf B~190}, 253 (1981); A.J. Heeger, S. Kivelson, J.R. Schrieffer, and
W.-P. Su,  Rev. Mod. Phys. {\bf 60}, 781 (1988); T.L. Ho,   J.R. Fulco, J.R.
Schrieffer, and F. Wilczek,  Phys. Rev. Lett. {\bf 52}, 1524 (1984).

\bibitem{Turok} N. Turok, Electroweak Baryogenesis, preprint
Imperial/TP/91-92/33

\bibitem{Diakonov} D. Diakonov, M. Polyakov, P. Sieber et al, Phys.
Rev. {\bf D~49}, 6864 (1995).

\bibitem{Kashiwaya} S. Kashiwaya, Y. Tanaka, M. Koyanagi, K. Kajimura,
"Bound states in superconductors", to be published in JJAP.

\bibitem{Tanaka} Y. Tanaka, S. Kashiwaya, Phys. Rev. Lett.
{\bf 74}, 3451 (1995).

\bibitem{vanOtterlo} A. van Otterlo, M. Feigel'man, V. Geshkenbein,
G. Blatter, preprint supr-con/9507004

\end{thebibliography}
\end{document}